\documentstyle[11pt,sao1,epsf]{article}

\newcommand{\km}{\,\mbox{km}\,\mbox{s}^{-1}}
\def\Ha{\hbox{H$_\alpha$~}}

\begin{document}
\large
\centerline{\LARGE \bf The structure and kinematics of  AGN}
\medskip
\centerline{\LARGE \bf with ionization cones.}

\bigskip

\centerline{ \bf A.V.Moiseev, V.L. Afanasiev, S.N. Dodonov}

\centerline{\it Special Astrophysical Observatory , Nizhnij Arkhyz,
Karachaevo-Cherkesia, 357147, Russia }

\medskip

\centerline{\bf V.V. Mustsevoi  and S.S. Khrapov}

\centerline{\it Volgograd State University, Volgograd, 400062, Russia}

\bigskip

\large

\centerline{\bf Abstract}
  Results of 2D  spectral  observations  for  4 Sy galaxies with ionization
cones  are  presented. Images in the [OIII] line,  velocity
fields  of  the  ionized  gas,  stellar  velocity and velocity dispersion
fields were obtained at the SAO RAS 6m telescope.
Non-circular  gas  motions and Z-shaped emission filaments could
be explained as helical waves  located in the ionization cone.
This waves are  generated by  the hydrodynamical instability due to the
velocity break between galactic ISM and outflowing matter from  the
AGN engine. The axis of cone lies  close to the  direction of
a nuclear mini bar.

\section{Introduction}
Many Seyfert galaxies   have a cone-like morphology  of Narrow
Line Regions (NLRs). In present work we discussed the structure of galaxies
with  following observational features:

\begin{itemize}
\item  an axis of cone lies along the  direction of a radio jet
(Wilson\&Tsevetanov,\cite{wt});

\item a multicomponent structure of the narrow emission lines
(e.g. [OIII]) shows that some systems of the gas clouds
with   velocity difference more than $100\div 1000 \km$
present in this galaxies.

\item emission filaments have a Z(S)-shape or/and arcs-like
("strands") structure;

\item a  large gradient of the line-of-sight gas velocities
presents in  Z-shaped spiral. Blueshifted and redshifted components
 of the emission lines has a different location
on  the sky-plane in  the  emission cone.

\end{itemize}
We consider results of  observations of the galaxies Mrk 3, Mrk 573,
NGC~3516 and NGC~5252. For each object some models for understanding
of a Z-shape morphology were proposed:
complex  of inclined and polar gaseous discs in NGC
5252 (Morse et al.,\cite{morse}) and in Mrk 3 (Afanasiev \&
Sil'chenko, \cite{afasil3}), bent bipolar outflow or
precessed  jet for NGC 3516 (Veilleux et al.,\cite{veil}).
Such  structures are unstable with short live-time, or
they should have very specific orientation relative to Earth observer, or
an unusual power is required (for jet precessing on scale $\sim3$
kpc). A new model of jet-cloud interaction for Mrk 3 (Rossi et
al.,\cite{rossi}) also cannot  describe a  Z-shaped emission pattern in NLR.

\section{The hypothesis.}

First, we assume that observable Z-shaped structures
of   NLRs are {\bf tridimensional helical  waves} in the ionizations cone.
These waves generated by  the hydrodynamical instability
due the velocity break between galactic ISM and a central
outflowing. A  collimated radio jet  associated with
direction of outflow and matches with the cone's axis.

Second, we considered the {\bf orientation of the cone}.
The Unified Model of AGN assumes the presence of circumnuclear disk
(torus) around the central engine which   collimates a jet
outflow and an ionized radiation. We propose to assume that torus orientation
associated with {\bf a particular directions} in a galactic body.
In an easy axisymmetrical case this is
direction of the rotation axis. All galaxies from our sample have a
triaxial structures  in central regions
(Mrk~573 and NGC~3516 are  double-barred galaxies,  Mrk~3 and
NGC~5252 have a probable central mini bar, see  section
3.1 and 3.4). In triaxial gravity potential, there are three planes where
stable orbits are possible:
this planes are lying along its axis.  Let us note that most
stable  orbits are located in the smaller dimensional plane, where
rotation matter has a minimal angular moment. The  gas in a galactic disk
decelerates by bar and accumulates to the torus. A resulting
radiation and outflowing cone will coincide with the angular momentum
vector of the torus (along   a major axis of the central bar
\footnote{Axon \& Robinson(\cite{axon}) and Vila-Vilaro (\cite{vila}) also
argued that orientation of the circumnuclear torus due the vertical inner ILR in a  secondary bar and
alignment with its direction.}).

\section{Observations}

The  observations with Interferometer Fabry-Perot  (IFP) for
galaxies Mrk~3, Mrk~573, NGC~3516 and NGC~5252
were obtained at the SAO
RAS 6m telescope in  1996-1998. Spectra of the [OIII]$\lambda5007$
emission line  were obtained  with  a spectral
resolution  $40-100 \km$ and  a spatial resolution 1\farcs 5-2\arcsec.
Mkn 3 and NGC 5252 were also observed with integral field spectrograph
MPFS in 1999-2000. A spectral range was
$\lambda4900-6100\AA$ and $\lambda4000-5200\AA$, a spectral resolution
$120-150\km$ and spatial data sampling  $1''/\mbox{lens}$ in
$16''\times15''$ field-of-view.
A cross-correlation method (Tonry\&Davis,\cite{td}) used
for stellar velocity and velocity dispersion measurements.
For Mrk 573 and NGC 3516 we used already published  MPFS-data
(Afanasiev et al.,\cite{afan573};
Afanasiev\&Vlasyuk,\cite{afanvvv}) with similar spectral and
spatial resolutions.

We fit the observable Z-shaped emission arms in  easy assumption of a
helical log-scale spiral on a cone's surface.
The axis of the cone was setting near the
galactic plane and projected on the sky-plane in alignment with radio
jet direction.  The  fitting results
are shown together with n [OIII] images and velocity fields (Fig. 1-4)

If  blueshifted and redshifted parts of the
helical spiral lie along a line-of-sight then a mutlicomponent
structure of emission line will be presented.This fact also was
taken in consideration.

\subsection{Mrk 3 (S0,  Seyfert 2)}

The  NLR has a bi-conical structure
which extended more than $8''$ ($2$~kpc), elongated in
$PA=114^o$ (Pogge\&De Robertic,\cite{pog}) and  shows a Z-shaped
structure on HST-images ( see Fig. 1a and Capetti
et al., \cite{capm3}). The analysis of the stellar velocity field
shows that  PA of a kinematical major axis shifted
on $7^o$ relative the photometrical $PA=23^o$ in $r<6''$.
This discrepancy relates with a
oval orbit distortion due to the large-scale bar with
$a\sim20''$. We fit the HST image by a 2D model (oblate
spheroid+disk+Ferrer's bar, Fig. 1b). A small-scale bar  may be also
presents in this galaxy (Capetti et al., \cite{capm3}) with  $a\le1''$
and  $PA\approx70^o$.

 The  profiles  of  the  [OIII] line in IFP-data demonstrate a
multicomponent structure (Fig.1c) in agreement with
long-slit data by Afansiev\&Sil'chenko(\cite{afasil3}).
The distance between  components is  $100-500 \km$.
A kinematical axis in the velocity field of main (brightest) component turns
away from the stellar axis more than $\sim80^o-120^o$.
A  velocity  gradient  along the helical spiral is presented in $r<7''$. The
second  component  of  the [OIII] profile also shows  non-circular motions
located  in  the  helical  structure only. This component  has  inverse
velocities in comparison with the main component of the emission line
profile (Fig. 1d). In $r<5''$ the
third   (low-brightness)  component of the [OIII] line  is   presented  and
corresponds to the stellar motions in the  large-scale bar.

From the helical fitting we obtained an estimation on the ionized cone
open angle $\theta _0=64^o$, and the inclination angle to the galaxy
plane $\delta i<10^o$.

\subsection{Mrk 573  (SAB0,  Seyfert 2)}

The NLR elongated in $PA\approx120^o$ and extended at $\sim13''$
($4.2$ kpc, Wilson\&Tsvetanov,\cite{wt}). This  shows an arc-like and
Z-shaped bi-conical structures on   ground-base and HST emission line
images (Fig. 2a and Ferruit et al.,\cite{ferr573}).
The optical image of  the galaxy can be fitted in double-barred
assumption  with   lengths of bars $5''$  and  $20''$  (Fig. 2b,  and
Moiseev,\cite{mois}).   The   secondary   bar   ($PA\approx115^o$)
distorts the stellar velocity field
(Fig. 2e)   in   $r<5''$.  The  [OIII]  line  in  $r<6''$  has  a
double-component profile (Fig. 2c) with separation $\pm200\km$ in
agreement with previous 2D spectroscopy study (Afanasiev et
al. \cite{afan573}). The main  component shows the strong
non-circular motions with $100-200 \km$ in amplitude. The second
component locates near the cone's axis and demonstrates an invert
sign of the velocities (Fig. 2d). Note that Ferruit et al.(\cite{ferr573})
carried out  a 2D  spectroscopy of the
galaxy and found  the strong red (blue) wings of the [OIII] line profile
on distance $2''-3''$ from the nucleus. In our  IFP data
these components are certainly decoupled (Fig. 2c).

From the helical fitting we obtained an estimation on cone open angle
$\theta_0=56^o$, the inclination angle to the galaxy
plane $\delta i\sim0^o$. The  axis of cones  align with a secondary
bar major axis.

\subsection{NGC 3516 (SB0,  Seyfert 1.5)}
The  image  in  [OIII]  line  shows  the  broad  cones  in
$PA=52^o$  (Fig. 3a). The NLR extended on $\sim38''$ ($6.2$ kpc) in NE
and  on $\sim14''$($2.3$ kpc) in SW  in agreement with
Miyaji  et  al.(\cite{miya}). The emission regions in $r<10''$ have
a Z-shaped morphology  also available in HST data (Ferruit
et.al \cite{ferr3516}).  We  fit  the  continuum image of NGC 3516
by a double-barred model, where bars have  sizes  $6''$ and
$22''$ (Fig. 3b). The  central mini-bar  has $PA\approx55^o$
and matches with  line-of-nodes of the  stellar velocity field from
Arribas et al.(\cite{arri}) and with our MPFS-data.
A  double-component  structure  of  the [OIII] profile (Fig. 3c and 3d), firstly
pointed out by Mulchaey et al.(\cite{mul}) is seen in $r<10''$. The
velocities of the emission components have a "blue-red" asymmetry
along the Z-shaped structure, in agreement with Afanasiev  et  al. observations
 at  the  4m  Mayall telescope. In $r>20''$ the
line-of-sight  velocities have a  gradient across the cone axis (along the
helical spiral arm).

From the helical fitting we obtained an estimation on cone open angle
$\theta_0=60^o$, the inclination angle to the galaxy
plane $\delta i<5^o$. The axis of cones align with a secondary
bar major axis.

\subsection{NGC 5252 (S0,  Seyfert 1.9)}
A beautiful bi-cone emission structure  extended in this galaxy
more than $50''$ ($\sim20$~kpc). The central Z-shaped arms and outer
arcs  are revealed  on  the [OIII] image (Fig. 4a. and Morse et
al.,\cite{morse}).  The major axis on J-band image (Alonso-Herrero
et  al.,\cite{alonso}) has $PA\approx18^o$ in agreement with
stellar rotation (Fig. 4a and 4d). We fitted the
J-image  by a  axisymmetrical  model  (disk+oblate  bulge), but the
structure  of  the  central isophotes is uncertain. The
asymmetry  in  stellar  velocity dispersion field and "boxy" isophotes
support   a  possible existence of a  mini bar ($r<5''$) 
in a high inclined disk.
The second component of the [OIII] line profile is presented in
$r<7''$  (Fig. 4a-c). Partly it has circular (as stellar)
rotation,  strong non-circular motions also available. The
velocity field of the bright [OIII] component has a very  strange structure
(Fig. 4b): isovelocities turned across the galactic major axis in the
central region and have an insignificant gradient at the outer part of the
galaxy (in the outer arcs). The gas line-of-sight velocities show large
disagreement $(100-400 \km)$  with the stellar one.

Note that  helical model matched with emission filaments (and its
velocities) only in $r<10-20''$. The outer emission arcs have a small
gradient of line-of-sight velocity and probably has a axis-symmetrical
structure.

\section{A numeric simulations}

We just started  sets of 2D and 3D  hydrodynamical simulations of
a conical jet with optically thin radiative cooling  in the field of
parabolic gravity potential. This simulations show  that  heavy damping of
all acoustical modes  and  little influence on unstable Kelvin-Helmholtz
surface modes are presented.
Waveguide-resonance internal gravity modes moving
relative to the jet matter from the source being always damped,
the growth of modes moving towards the source is faster
then in adiabatic jet.  Our analysis shows that
surface Kelvin-Helmholtz modes and waveguide-resonance internal
gravity body  modes will be  most effective at producing shock waves
outside from the central outflow (radio jet).
 Moreover, these shocks create
a cone with rather large opening angle than initial jet.
The intensity of radiative cooling in these shocks {\bf provide
 heating of the ambient medium and forms a cone of ionization.}

Our simulation shows that  axisymmetrical waves are grows slower than
helical waves. Therefore these waves can be observed at the {\bf outer regions}.
An amplitude of the helical modes decrease with radius (unlike from
axisymmetrical modes). And nonlinear superposition of helical and
axisymmetrical waves can give the structure which observed in
the NGC~5252. Moreover,  simulation shows that line-of-sight
gas velocities will be increased with a distance from center,
in good agreement with observations (section 3).

The Fig.4e and Fig.4f. reproduce projections of the ionized gas luminosity
on the sky plane which were obtained from the 2D and 3D simulations.
We can see axisymmetrical waves ("rings") and helical waves around
the central outflow.
These first results  support our scenario of formation of the
Z-shaped and arc-like emission structures in  Seyfert galaxies.

{\bf We postulate  that  model of NLR which includs the bar, radio jet,
ionization cone and helical waves in the cone may be constructed
from easy assumption, without special limitations on the origin of  AGN
central engine.}

\acknowledgements{
We would like to thank astronomers of the SAO -- Alexander Burenkov,
Irina Kostiuk, Alla Shapovalova and Valery Vlasyuk
for  assisting at the 6m telescope, Sergej Drabek and Eugene Gerasimenko for
technical supporting of the observations and 6m telescope committee
for allocating observing time.
Also we are grateful to Viktor Levi for helpful discussions and
to  to Almudena Alonso-Herrero  who provided the J-band image of NGC 5252.
}

\newpage

\begin{figure}
\centerline{\epsfbox{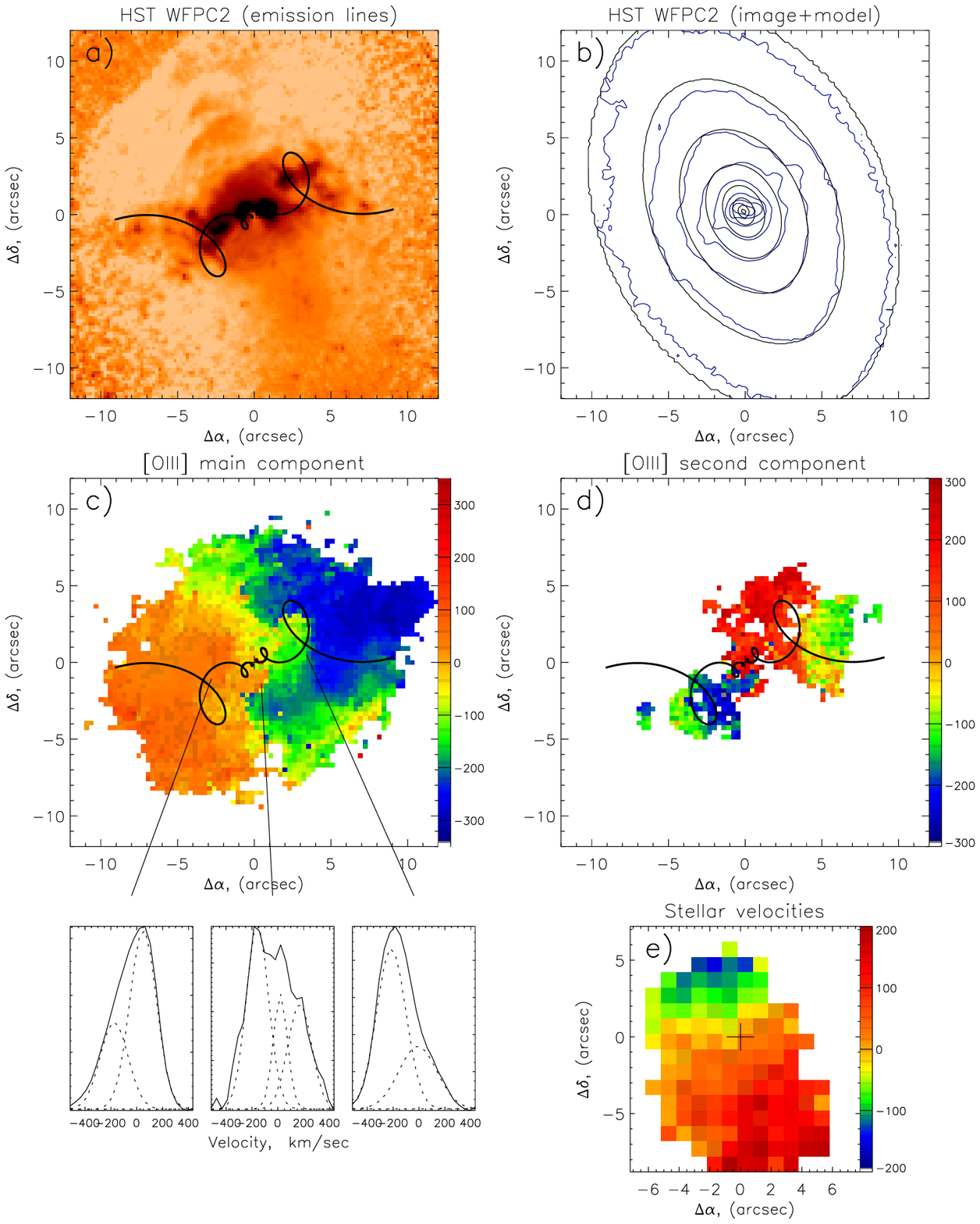}}
\caption{
 {\bf  Mrk 3}.
 (a)  Residual image: 2D-model of the continuum brightness
was subtracted from the WFPC2 image in the filter F606 ( HST data
 archive).
The [OIII]$\lambda 5007$, [OI]$\lambda 6300$ and
\Ha  emission  lines  has contribution in the residual  brightness.
(b) (blue) -- HST WFP2 image;
(black) --  isophotes of the 2D model (oblate
bulge+disc+ Ferrer's bar).
(c)  Velocity field of the main component of
the [OIII] emission line. A color velocity scale -- in $\km$~
(in relative to  system velocity $V_{SYS}=4000\km$).
Examples of the [OIII] emission line profiles  and
gauss-decomposition of components show at the bottom panels.
(d) Velocity field of the second  component
of [OIII].
(e) Stellar velocity field.
}
\end{figure}

\begin{figure}

\centerline{\epsfbox{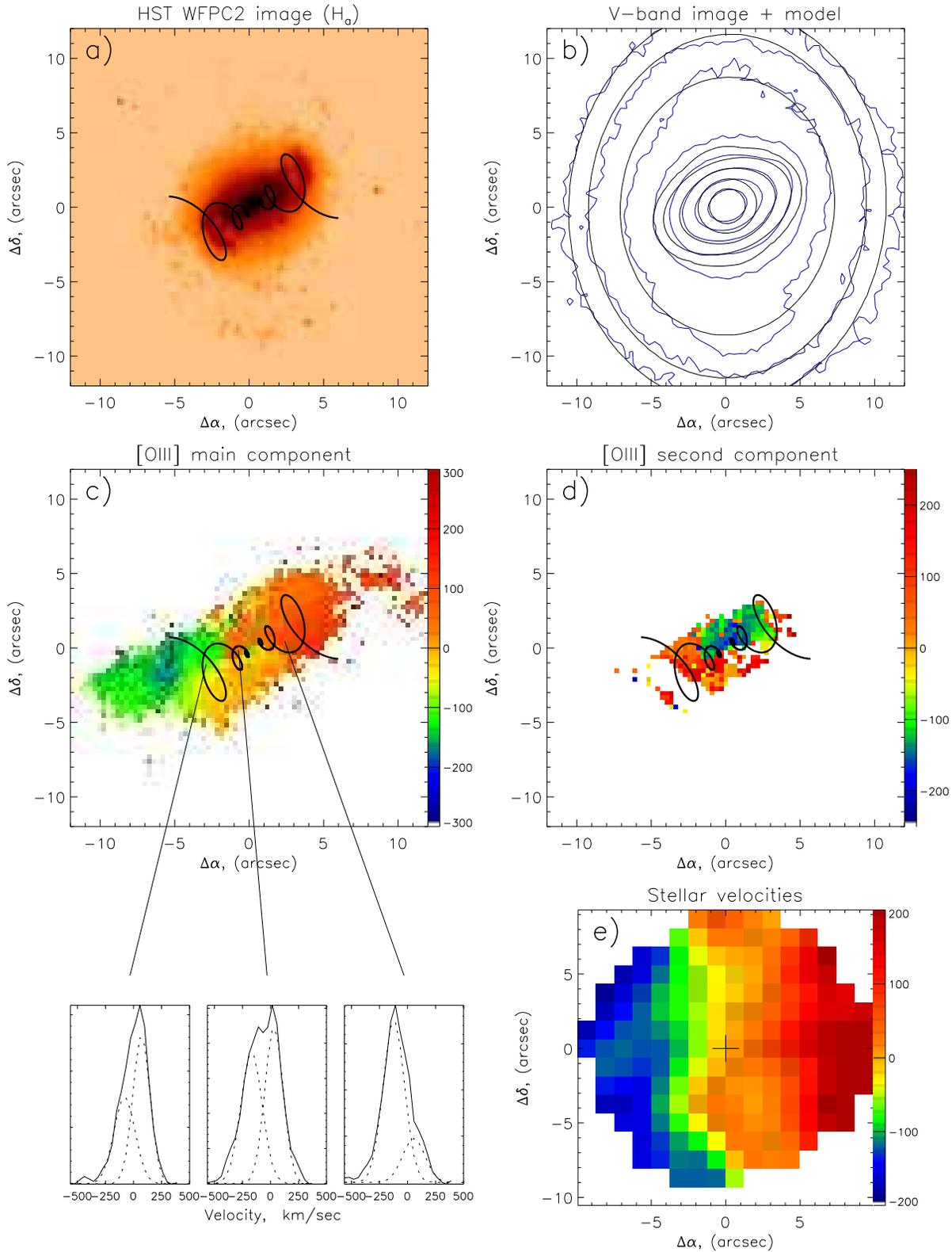}}
\caption{  {\bf  Mrk 573}
(a) -- HST WFPC2 image in the \Ha emission line (filter FR680N, HST data
archive).
(b) (blue) -- isophotes  of the V-band image (SAO RAS 1m
telescope), (black) -- isophotes of the 2D model (bulge+disc+two
Ferrer's bars).
(c)  Velocity field of the main component of
[OIII] ($V_{SYS}=5160\km$).
(d)  Velocity field of the second  component of [OIII].
(e) Stellar velocity field.
}
\end{figure}

\begin{figure}
\epsfxsize=19 cm
\centerline{\epsfbox{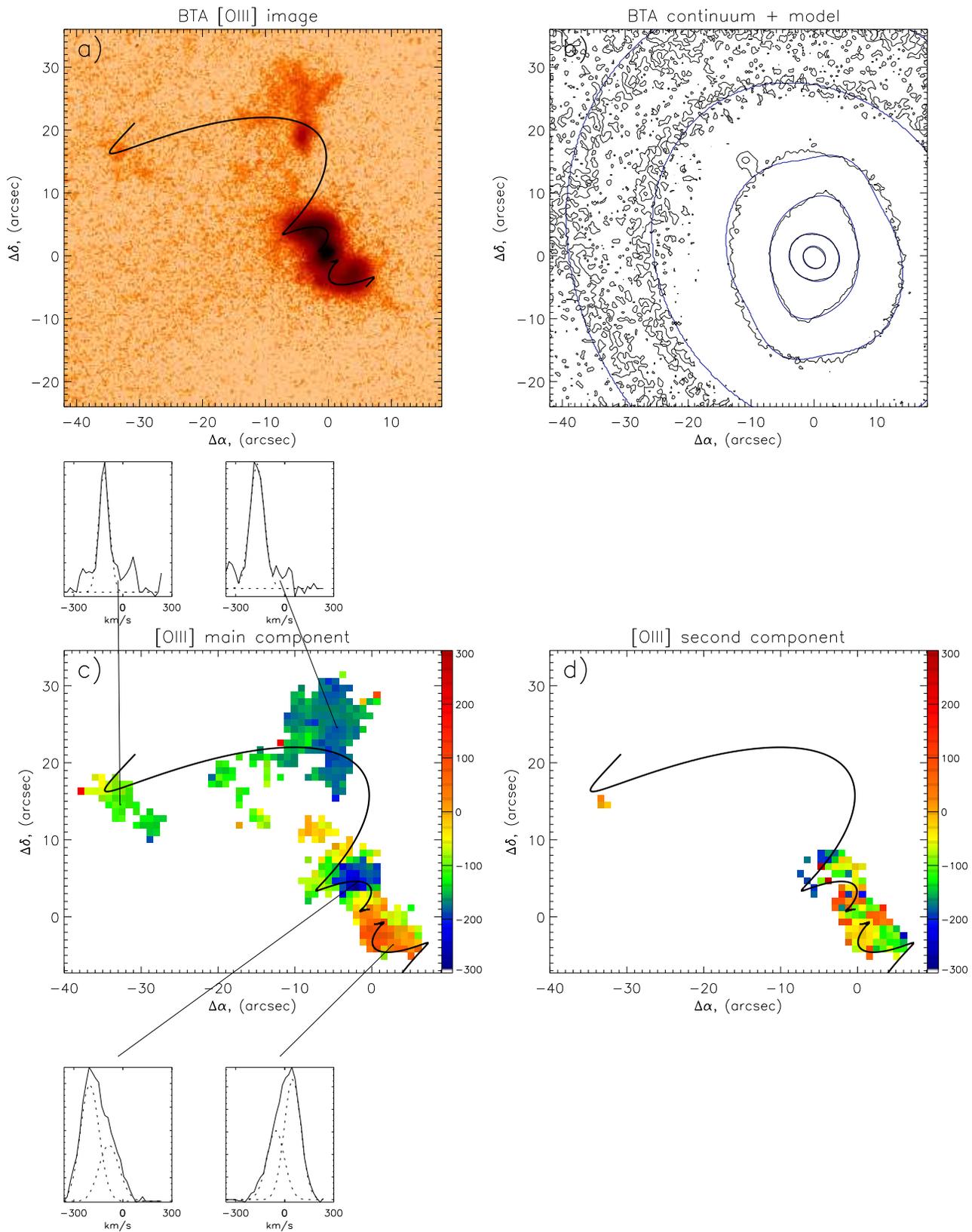}}
\caption{
{\bf  NGC 3516}
(a) Image in the [OIII] emission line (6m telescope).
(b) (blue) -- continuum isophotes near [OIII];
 (black) --  isophotes of the 2D model (bulge+disc+two Ferrer's bars).
(c) Velocity field of the main component of
 [OIII] ($V_{SYS}=2560 \km$).
(d) Velocity field  of the second component of [OIII].
}
\end{figure}

\begin{figure}
\centerline{\epsfbox{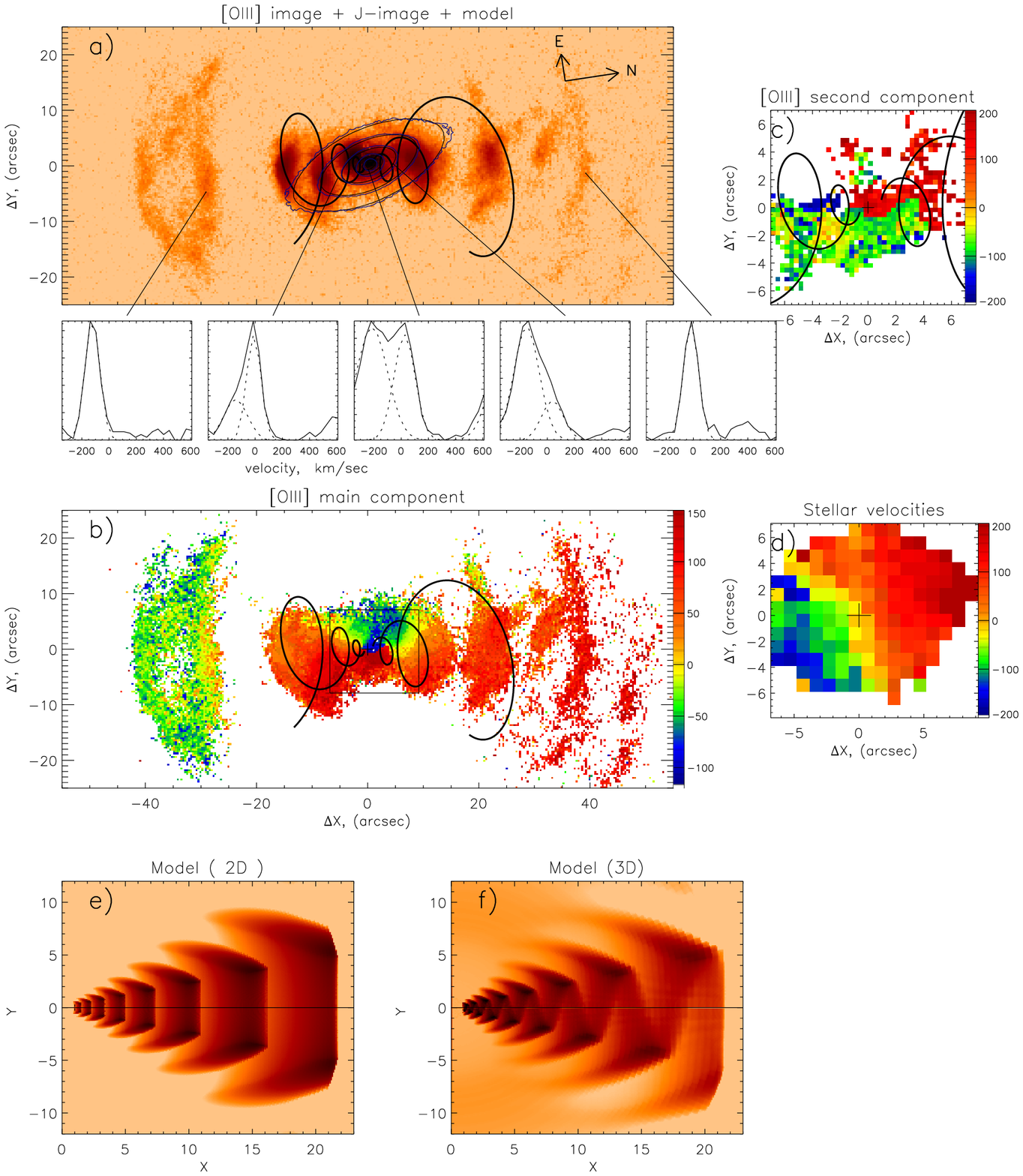}}
\caption{
{\bf  NGC 5252}
(a) Image in the [OIII] emission line (IFP-data).
Blue -- isophotes of the J-band image,
black --  isophotes of the 2D model (bulge+disc)
(b) Velocity field  of  the main component of [OIII]
($V_{SYS}=6880 \km$).
(c) Velocity field of the  second  component
of [OIII].
(d) Stellar velocity field.
{\bf Results of the simulations:} projection of the model luminosity
on the sky-plane. (e) axisymmetrical waves ("rings").
(f) non-axisymmetrical waves ("helicon").
}
\end{figure}

\begin{thebibliography}{}

\bibitem[1995]{afan573} Afanasiev, V.L., Burenkov, A.N., Shapovalova,
A.I., Vlasyuk, V.V.,1996, in IAU Coll.157,Vol.91, 218

\bibitem[1995]{afanvvv} Afanasiev, V.L., Vlasyuk, V.V.,1995, in IAU Coll.149,Vol. 71, 266

\bibitem[1990]{afasil3} Afanasiev, V.L. \& Sil'chenko, O.K., 1991,Bull.
SAO,33,104

\bibitem[1998]{alonso}
Alonso-Herrero, A., Simpson, C., Ward, M.J., Wilson, A.S.,1998,ApJ,495,196

\bibitem[1997]{arri}
Arribas, S.,Mediavilla, E.,  Garsia-Lorenzo, B., Del Burgo, C., 1997,ApJ,490,227

\bibitem[1996]{axon}
Axon, D.J. \& Robinson, A., 1996, in Nobel Symposium 98,223


\bibitem[1995]{capm3}
Capetti, A., Macchetto, F., Axon, D.J., et al.,1995,ApJ,448,600

\bibitem[1999]{ferr573}
Ferruit, P., Wilson A.S., Falcke, H., et al.,1999, MNRAS,309,1

\bibitem[1998]{ferr3516}
Ferruit, P., Wilson, A.S., Mulchaey, J.S.,1998, ApJ,509,646

\bibitem[1992]{miya}
Miyaji, T., Wilson, A.S., Perez-Fournon, I.,1992,ApJ,385,137

\bibitem[1998]{mois}
Moiseev, A.V., 1998, SAO RAS preprint, N 134/1

\bibitem[1998]{morse}
Morse, J. A., Cecil, G., Wilson, A.S., Tsvetanov, Z.I.,1998,ApJ,505,159

\bibitem[1992]{mul}
Mulchaey, J.S., Tsvetanov, Z., Wilson, A.S., Perez-Fournon, I.,1992,ApJ,394,91

\bibitem[1993]{pog}
Pogge, R.W. \& De Robertic, M.M., 1993,ApJ,404,563

\bibitem[2000]{rossi}
Rossi, P., Capetti, A., Bodo, G., et al.,2000,A\&A,356,73

\bibitem[1979]{td} Tonry, J. \& Davis, M.,1979,AJ,84,1511

\bibitem[1993]{veil}
Veilleux, S., Tully, R.B., Bland-Hawtorn, J., 1993,AJ,105,1318

\bibitem[1995]{vila}
Vila-Vilaro, B., Robinson, A., Perez, E., et al.,1995,A\&A,302,58

\bibitem[1994]{wt} Wilson, A.S. \& Tsvetanov, Z.I.,1994,AJ,107,1227
\end{thebibliography}
\end{document}